# THE QUANTUM COHERENCE AS A PHASE PROPERTY OF THE WAVE FUNCTION

I. G. KOPRINKOV

Department of Applied Physics, Technical University of Sofia,
8 Kl. Ochridski Blvd., Sofia 1000, Bulgaria
*E-mail*: igk@tu-sofia.bg

*Abstract*. *The quantum coherence as a phase property of the most fundamental quantum mechanical entity, the wave function, is explained for the first time. Phase-sensitive nonadiabatic dressed states, arising from the interaction of a quantum system and a coherent external electromagnetic field, are used in these considerations. Two types of phase correlations in the multilevel phase-sensitive nonadiabatic dressed states are found: a rapidly changing phase correlation between the real and the virtual components and a stationary phase correlation between different virtual components of these states.*

## 1. INTRODUCTION

The quantum coherence is one of the most remarkable properties of the quantum phenomena, which sets the border between the quantum and classical behavior of the matter. As such, the quantum coherence becomes closely related to other fundamental quantum phenomena as quantum superposition, quantum interference, quantum entanglement, quantum decoherence, etc. [1-5]. In addition, the quantum coherence is at the heart of application of quantum physics in the quantum information science [6], quantum cryptography [7], etc. A growing interest toward a resource theory of quantum coherence has been demonstrated in relation to its technological applications [8].

The quantum coherence encompasses different concepts in the different fields of quantum physics [1]. In quantum mechanics, the simultaneous and coherent superposition of quantum states can be considered on the basis of the quantum superposition principle, which formally holds due to the linearity of the Schrödinger equation. The quantum coherence is a signature of definite correlations between the states of the quantum system and represents the ability of the quantum states to interfere. Such correlations are usually described by the off diagonal elements of the density matrix. The coherence of quantized bosonic/optical fields is considered as existence of definite correlations upon photon detection coincidence, *e.g.*, the Hanbury Brown and Twiss effect [9]. These are additional intensity correlations to the ordinary optical phase correlations. The coherence of quantized bosonic fields can be treated within the Glauber–Sudarshan coherent states [10, 11], which are eigenstates of the annihilation operator and represent the minimal uncertainty states.

In this work, the quantum coherence is explained for the first time as a phase property of the most fundamental quantum mechanical entity, *i.e.*, the wave function. This approach rests on the earlier proven causal relation between the dynamical phase of wave function and the physical reality, called material phase (MP) causality, based on a large number of theoretical and experimental evidences [12]. The dynamics-statistical interpretation of quantum mechanics has been introduced on this ground [12]. The dynamics-statistical interpretation of quantum mechanics is the only interpretation that recognizes the MP causality while preserving at the same time the epistemological meaning of the wave function.

The quantum coherence is closely related to the ability of quantum states to interfere, which imposes the requirement that the quantum states must be superimposed simultaneously, or, at least, to have a non-zero time of coexistence of the states within the quantum system. Both cases of superposition will be called, for short, *simultaneous* although, when necessary, the difference between them will be specified. In a previous work [13], we have shown that some of the most widely used quantum states, *e.g.*, the eigenstates of stationary or adiabatic Hamiltonians, cannot be superimposed simultaneously and therefore physically cannot interfere. A particular case of stationary states are the bare states (BS), the states of a completely isolated, closed, quantum system. In searching for a simultaneous superposition of quantum states, we rely on the physical mechanism of creation of the states. Based on such an approach, we have reformulated the quantum superposition principle. This, in addition, allows us to explain the collapse of wave function and the quantum measurement problem as natural processes based solely on physical processes including in the Hamiltonian of the quantum system, *e.g.*, coherent interactions with external coherent electromagnetic field and incoherent interactions due to the environment. In particular, we showed [13] that the real and virtual components of nonadiabatic dressed states (NADS) [14] satisfy the requirement of simultaneity. To completely account for the quantum coherence, all relevant phase factors, even the constant ones, must be taken into account and, therefore, phase-sensitive nonadiabatic dressed states (PSNADS) [15, 16], an extended version of NADS, will be used in the present considerations. Two types of phase correlations are found between the different components of the PSNADS (multi-level case): (*i*) a rapidly changing phase correlation between the real and virtual components within given PSNADS and (*ii*) a stationary phase correlation between the different virtual components within given PSNADS.

## 2. QUANTUM COHERENCE AND PHASE-SENSITIVE NONADIABATIC DRESSED STATES

To reveal the MP dynamics, the quantum system must be involved in definite physical processes, the most important of which are interactions with an external coherent electromagnetic field and incoherent interaction with the environment. The rigorous solution of this problem naturally leads to PSNADS, which include phase contributions from the matter (MP) and the field, and nonadiabatic contributions from the field (time derivative of the field amplitude and phase) and the environment (damping rates).

### 2.1 PHASE PROPERTIES OF TWO-LEVEL PHASE-SENSITIVE NONADIABATIC DRESSED STATES

For the two-level case, the PSNADS arise from a nonadiabatic and non-perturbative closed form solution of the Schrödinger equation for a quantum system interacting with an external electromagnetic field (whose amplitude and phase are arbitrary functions of time, subject only to a generalized adiabatic condition [14, 15]) and environment (presented by damping rates of the states) described the Hamiltonian (in standard notations) [15]:

$$\hat{H} = \sum_{j=1}^{2} \hbar\omega_j |j\rangle\langle j| - \mu E(t)\big(|1\rangle\langle 2| + h.c.\big) - i\hbar\sum_{j=1}^{2}\big(\gamma_j/2\big)|j\rangle\langle j| \quad , \tag{1}$$

All relevant phase contributions are taken into account, including the initial constant phases $\varphi_g$ and $\varphi_e$ of the initial ground $|1\rangle \equiv |g\rangle$ and excited $|2\rangle \equiv |e\rangle$ bare states (BS), respectively, from which the PSNADS originate due to the above mentioned interactions. The total MP of the real (index R) and virtual (index V) components of ground (index G) and excited (index E) PSNADS, Fig.1, *e.g.*, at ground state initial conditions, are:

$$\Phi_{G,R} = \varphi_g + \int_0^t \tilde{\omega}_{G,R}\, dt' \tag{2a}$$

$$\Phi_{G,V} = \Phi_{G,\mathrm{R}} + \Phi_F = \varphi_g + \varphi(t) + \int_0^t (\tilde{\omega}_{G,R} + \omega) dt' \tag{2b}$$

$$\Phi_{E,R} = \Phi_{\mathrm{G,V}} + \Phi_{NAD} = \varphi_g + \varphi(t) + \int_0^t \tilde{\omega}_{E,R}\, dt' \tag{2c}$$

$$\Phi_{E,V} = \Phi_{E,\mathrm{R}} - \Phi_F = \varphi_g + \int_0^t (\tilde{\omega}_{E,R} - \omega) dt' \quad , \tag{2d}$$

where $\tilde{\omega}_{G,R}$ and $\tilde{\omega}_{E,R}$ are the Bohr frequencies/energies of real components of ground and excited PSNADS (Stark shifted and modified by the nonadiabatic factors from the field and damping rates $\gamma_g$ and $\gamma_e$), $\Phi_F(t) = \varphi(t) + \omega t$ is the total field phase, $\omega$ is carrier frequency of the field, $\Delta\tilde{\omega}_{NAD} = \tilde{\omega}_{E,R} - \tilde{\omega}_{G,R} - \omega$ is nonadiabatic frequency detuning, and $\Phi_{NAD} = \int_0^t \Delta\tilde{\omega}_{NAD} dt'$ is the phase contribution due to the nonadiabatic transition between ground and excited PSNADS, more particularly, $|\tilde{G}_V\rangle \to |\tilde{E}_R\rangle$ and $|\tilde{E}_V\rangle \to |\tilde{G}_R\rangle$ transitions, Fig.1. Similar equations follow at excited state initial conditions. The detailed study of the MP evolution leads to the following conclusion: *the MP behaves as an additive dynamical quantity that causally follows the physical processes and the initial conditions* [15].

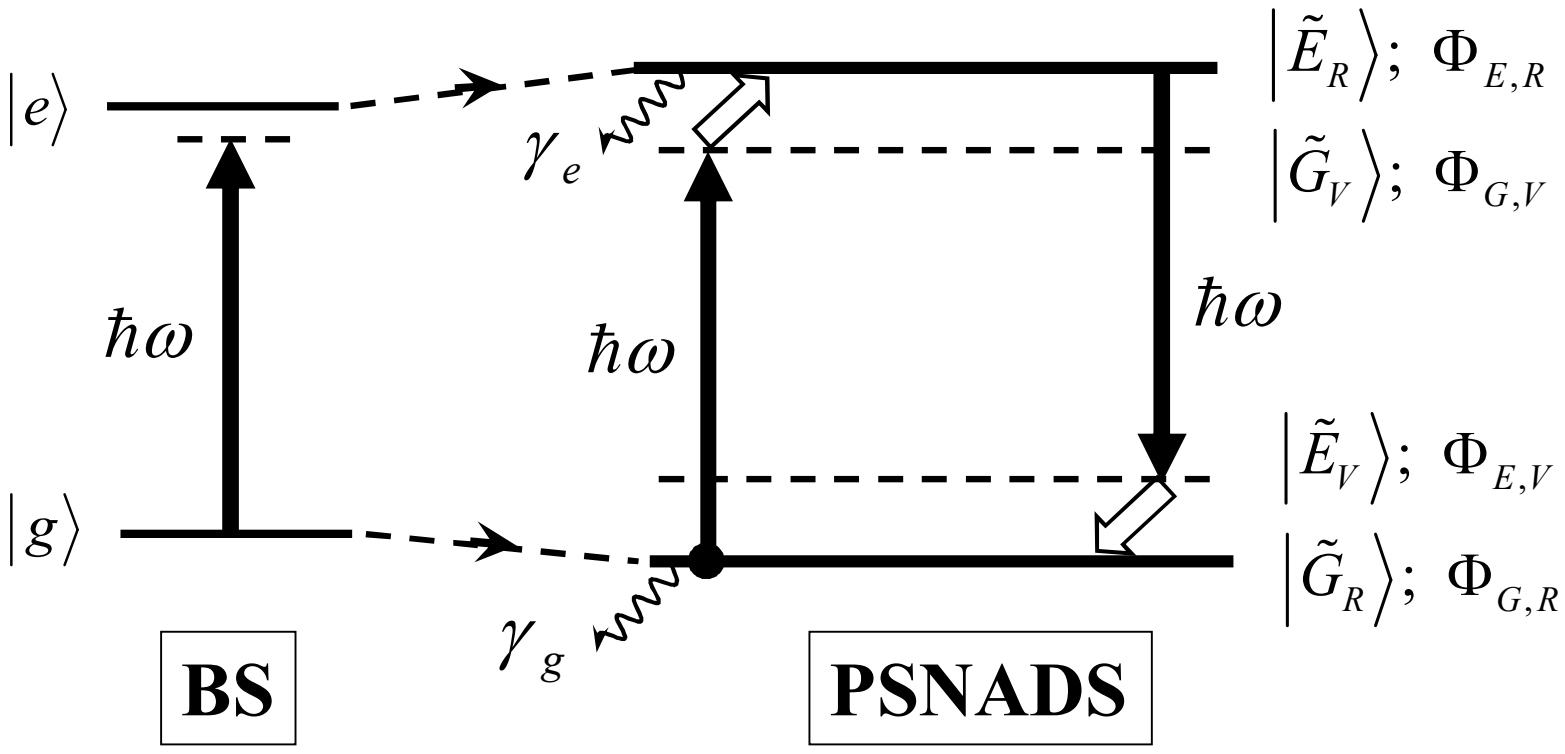

Fig.1. Energy level diagram of the quantum system in BS and PSNADS picture. Material phase of the real and virtual components of two-level PSNADS.

## 2.2 PHASE PROPERTIES OF MULTI-LEVEL PHASE-SENSITIVE NONADIABATIC DRESSED STATES

Closed form solution of the PSNADS is known only for the two-level case. In order to better understand the quantum coherence, multilevel PSNADS have to be considered. The multilevel PSNADS will be obtained extrapolating the two-level case in the following way. For, *e.g.*, ground PSNADS (at ground state initial conditions), if we take the original ground BS $|g\rangle$ and repeat the procedure of derivation of PSNADS, as in [15], in a consecutive order with every one of the (electric-dipole allowed) excited BS $|e\rangle_1$, $|e\rangle_2$, $|e\rangle_3$..., the result will be generally same, as in the two-level PSNADS, with the difference that for each excited BS (due to the different detuning $\Delta\omega_i$ from the exact resonances, Fig.2, and different dipole moment matrix elements between the states), the superposition coefficients (strengths) of the respective virtual states, denoted by $C_{Vi}$, $i=1,2,3..$, (instead of $SIN(\theta/2)$, as for the two-level case [15]), and the dynamic Stark shift of the states will be different. The smaller is the detuning $\Delta\omega_i$ of the photon energy from the exact resonance with given excited BS $|e\rangle_i$ and bigger is the dipole matrix element between the excited BS $|e\rangle_i$ and the ground BS $|g\rangle$, the bigger is the contribution of this excited BS $|e\rangle_i$ to the created virtual component $|G_V\rangle_i$, *i.e.*, bigger is the superposition coefficient $C_{Vi}$ of this virtual component within the multilevel

PSNADS, and vice versa. At the same time, the ground BS $|g\rangle$ of Bohr frequency/energy $\omega_g$ evolves toward a single (non-degenerate case) real component $|G_R\rangle$ of ground PSNADS of a well-defined Bohr frequency/energy $\tilde{\omega}_{G,R}$, whose value (position on the energy scale) is different for different $|g\rangle - |e\rangle_i$ pairs, due to the different Stark shift as a result of different frequency detuning $\Delta\omega_i$ and different dipole matrix element between different pair of states. The Bohr frequency/energy of each virtual component $|G_V\rangle_i$, created within given pair of states between $|g\rangle$ and each of the excited BS $|e\rangle_i$, is always equal to the Bohr frequency/energy of the (single) real ground state $\tilde{\omega}_G$ plus one photon frequency/energy $\omega$, $\tilde{\omega}_{G,V} = \tilde{\omega}_{G,R} + \omega$, as in the two-level case [15]. If we now allow all these excited BS to act simultaneously, as it is in reality, each excited BS $|e\rangle_1$, $|e\rangle_2$, $|e\rangle_3$..., while evolving toward respective real component $|E_R\rangle_1$, $|E_R\rangle_2$, $|E_R\rangle_3$..., will contribute to the creation of a respective virtual component $|G_V\rangle_1$, $|G_V\rangle_2$, $|G_V\rangle_3$... of ground PSNADS, Fig.2. In this process, the ground BS $|g\rangle$ of Bohr frequency/energy $\omega_g$ again (as in the two-level cases) evolves toward *a single only* (non-degenerate case) real component $|G_R\rangle$ of the ground PSNADS of a well-defined Bohr frequency/energy $\tilde{\omega}_{G,R}$. The carrier frequency (photon energy) $\omega$ of the field also keeps a well-defined value (at absence of nonlinear processes).

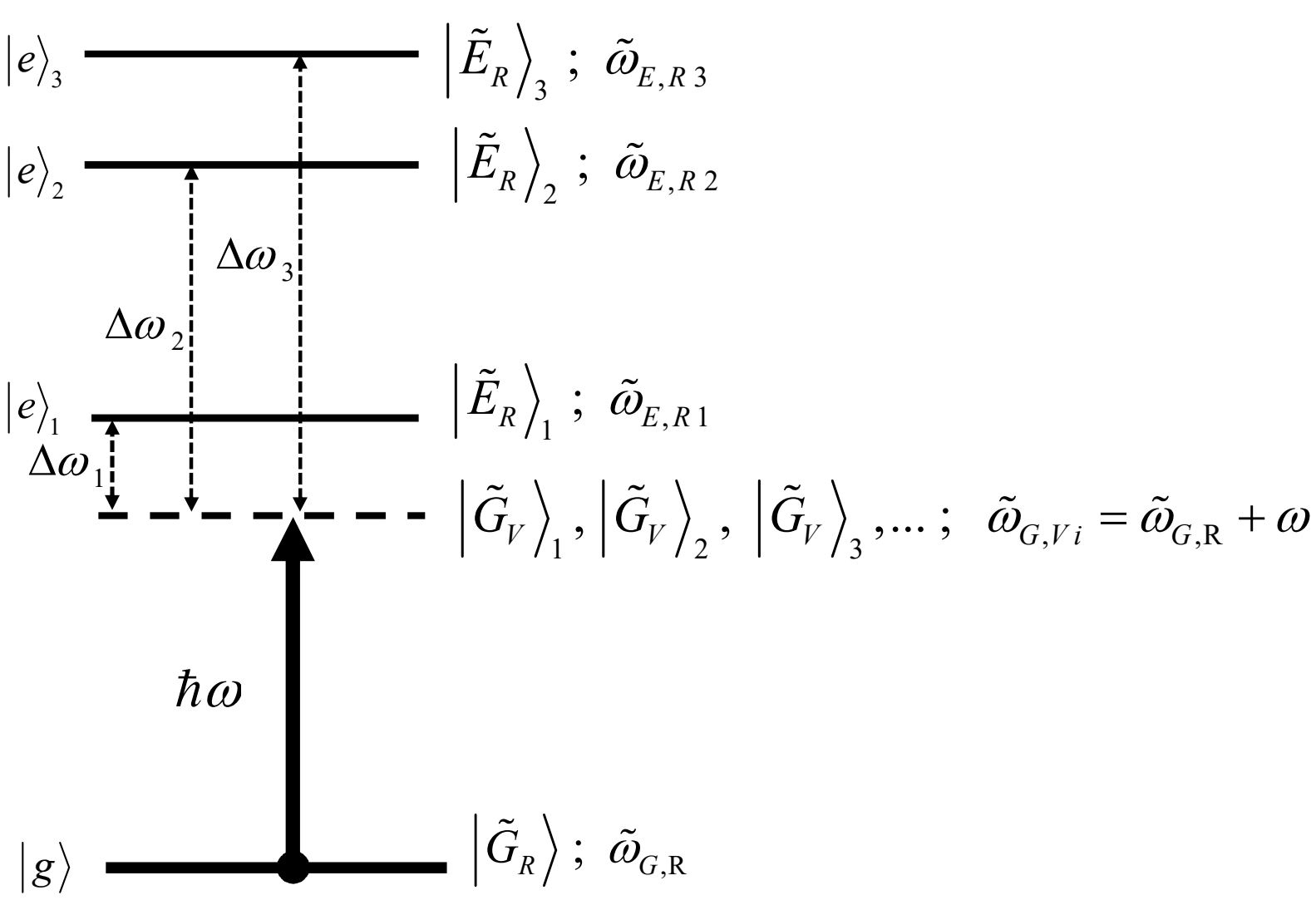

Fig.2. Multi-level PSNADS. The initial BS are shown at the left hand side of the figure. The real components (full lines) and the virtual components (broken line) of ground PSNADS together with the Bohr frequencies/energies of the respective components are shown at the right of the figure.

This means that all virtual components $|G_V\rangle_1$, $|G_V\rangle_2$, $|G_V\rangle_3$ …, have equal Bohr frequencies/energies, Fig.2:

$$\tilde{\omega}_{G,\mathrm{V},i} = \tilde{\omega}_{G,\mathrm{R}} + \omega = \tilde{\omega}_{G,\mathrm{V}} = const \quad , \quad i = 1, 2, 3.. \tag{3}$$

Same Bohr frequency/energy of all virtual components has crucial importance for the coherence of the virtual components of multilevel PSNADS in the very same way as same optical frequency (photon energy) has crucial importance for the coherence of the optical fields.

Based on the above considerations, the following conclusions about the structure and the MP of the multi-level PSNADS can be done. Each PSNADS consists of a single real component and a number of virtual components, equal to the number of the initial electric-dipole allowed BS involved in the creation of PSNADS. Thus, the structure of, *e.g.*, multi-level ground PSNADS, is:

$$|G\rangle = C_R\,|G_R\rangle + \sum\nolimits_i C_{Vi}\,|G_V\rangle_i \tag{4}$$

The superposition coefficients (strengths) of the real and virtual components, $C_R$ and $C_{Vi}$ respectively, of the multi-level case stand for the respective superposition coefficients (strengths), $COS(\theta/2)$ and $SIN(\theta/2)$, of the two-level case [15]. Similar structure will also have multi-level excited PSNADS, but, beside of a single real component, it consists of "upward" and "downward" virtual components due to upward and downward excitations of the quantum system by the field. The strength, $SIN(\theta/2)$, of the virtual component is zero at zero-field and nonzero at nonzero-field and increases with the field strength [14, 15]. In contrast to the real component, which naturally exists in the quantum system within the lifetime of the state, the virtual component cannot exist independently from the respective real component, from which it originates, and the forcing electromagnetic field, which is responsible for the creation of the virtual component. Extrapolating these properties toward the multi-level case, similar behavior for the superposition coefficients $C_{Vi}$ of the virtual components for the multilevel PSNADS is expected because same physics is responsible for the formation of virtual components of the two-level and the multilevel case: the superposition coefficients $C_{Vi}$ are zero at zero-field and nonzero at nonzero-field. This means that *all virtual components of given PSNADS appear simultaneously and synchronously, i.e., coherently* (see below), with the field from the respective real component with switching the field on and disappear with switching the field off. On the other hand, the mechanism of formation of virtual components from the respective real component show that *the superposition of real and virtual components of given PSNADS is simultaneous within the overlap time of the real component lifetime and the time of action of the field that creates the virtual components*. It can be referred to *the second type of simultaneity, i.e., a non-zero* *time of coexistence of the states*, as it is specified above. The existence of simultaneity in superposition between the real and virtual components of PSNADS is in sharp contrast to the properties of the most widely used BS, which, as has been shown earlier [13], cannot be superimposed simultaneously and coherently.

The Bohr frequency is the time rate of MP acquisition of the state while the MP is time integral over the respective Bohr frequency, *e.g.*, Eqs.(2). As the Bohr frequency (the integrand) $\tilde{\omega}_{G,\mathrm{V,i}} = \tilde{\omega}_{G,\mathrm{V}}$ of all virtual components $|G_V\rangle_i$ is same, Eq.(3), the total MP of the different virtual components, *i.e.*, time integrals over $\tilde{\omega}_{G,\mathrm{V,i}}$, may differ only by a constant phase shift $\varphi_{G,\mathrm{V},i} = const$ , $i = 1, 2, 3..$ , for each virtual component $|G_V\rangle_i$ in the process of its creation. The total phases of the real and the virtual components of the multi-level ground PSNADS, after modification of Eqs. (2a) and (2b), *i.e.*, adding a constant phase shift $\varphi_{G,\mathrm{V},i}$ to each virtual component, are:

$$\Phi_{G,R} = \varphi_g + \int_0^t \tilde{\omega}_{G,R} dt' \tag{5a}$$

$$\Phi_{G,V,i} = \varphi_{G,\mathrm{V},i} + \varphi_g + \varphi(t) + \int_0^t (\tilde{\omega}_{G,\mathrm{R}} + \omega) dt' \quad , \tag{5b}$$

where $i = 1, 2, 3...$ is the number of virtual state involved in the multi-level ground PSNADS.

The electromagnetic/optical fields that create the PSNADS are, in general, non-monochromatic and the virtual components of the PSNADS will have a nonzero spectral/energy width. In this case, the superposition of virtual components may lead to formation of localized material wave-packets inside the quantum system. This allows tracing the internal dynamics of the quantum system (atoms, molecules, etc.), especially if the virtual components, and PSNADS as a whole, are created by ultrashort laser pulses. It is important to note that the virtual components of the PSNADS are real physical states, but not simply a mathematical construct, due to a number of experimental and theoretical evidences [14].

### 3. THE QUANTUM COHERENCE WITHIN PHASE-SENSITIVE NONADIABATIC DRESSED STATES

The spectral/phase properties of the PSNADS, *i.e.*, same Bohr frequency/energy and a constant phase shift of the different virtual components, have a crucial importance for the quantum coherence in terms of quantum states as well as for the interference of quantum states. Another important point for the quantum interference is the simultaneity in the superposition of the quantum states, which also takes place between the real and virtual components of the PSNADS.

Seeking for quantum coherence of the states, the phase differences between (*i*) the virtual and the real components, Eq.(6a), and (*ii*) between the different virtual components of given PSNADS, Eq.(6b), will be considered based on Eqs.(5):

$$\Phi_{G,V,i} - \Phi_{G,R} = \varphi_{G,\mathrm{V},i} + \varphi(t) + \omega t \tag{6a}$$

$$\Phi_{G,V,i} - \Phi_{G,V,j} = \varphi_{G,\mathrm{V},i} - \varphi_{G,\mathrm{V},j} = const \quad , \tag{6b}$$

where $i, j = 1, 2, 3 \ldots$, and $i \neq j$ .

Two types of phase correlations can be distinguished from Eqs.(6):

(*i*): "*fast"* correlation or "*low coherence*", Eq.(6a).

(*ii*) "*slow*" correlation or "*high coherence*", Eq.(6b).

As can be seen from Eq.(6a), a well-defined regular phase difference exists between the phases of the virtual and the real components of the PSNADS, although it rapidly changes in time due to the large phase acquisition term $\omega t$, where $\omega$, beside of the carrier frequency of the field, represents the frequency/energy difference between the virtual and the real components, $|G_V\rangle_i$ and $|G_R\rangle$, respectively, Fig.2. Interference of this type is hard to be observed experimentally because the superposition of $|G_V\rangle_i$ and $|G_R\rangle$ components changes in time with a huge rate in the order of the optical frequency $\omega$. Therefore, it will be considered as a "hidden" coherence. On the other hand, Eq.(6b) shows an existence of a constant phase difference between any two virtual components $|G_V\rangle_i$ without rapidly oscillating term because all virtual components have same Bohr frequency/energy $\tilde{\omega}_{G,\mathrm{V}i} = \tilde{\omega}_{G,\mathrm{R}} + \omega = \tilde{\omega}_{G,\mathrm{V}}$. This type of coherence leads to a stable and a well (but not easy) observable quantum interference distribution between the different virtual components $|G_V\rangle_i$ of the PSNADS. Therefore, it will be considered as a "high coherence". Part of these properties become clear even within the strict two-level solution, but the complete picture of the phase coherence can be revealed within the multilevel PSNADS. Here, we focus our attention on the coherence of ground PSNADS while similar coherence behavior can be found if the process develops at excited state initial conditions.

The following *definition of quantum coherence* will be introduced based on the properties of "*high coherence*": *two or more (simultaneously superimposed) quantum states are coherent if they have same energy (Bohr frequency) and a constant difference between phases of these states*. The present approach allows generalizing the notion of "coherence" putting on the same footing the quantum coherence and the ordinary coherence of optical fields.

In agreement with adiabatic theorem of quantum mechanics, transitions between different states result from nonadiabatic factors acting on the quantum system. The nonadiabatic factors have, in general, a stochastic nature due to, at least, the zero-point vacuum fluctuations, which act universally on any quantum system and cannot be physically removed, even in principle. Hence, transitions between different PSNADS lead to stochastic phase contributions and this destroy the quantum coherence. Strictly speaking, *quantum coherence may exist within given PSNADS*, or other kind of quantum states created under the action of coherent forcing physical factor.

## 4. CONCLUSIONS

The quantum coherence as a phase property of the wave function is explained for the first time. Two types of phase coherence between virtual and real components as well as between different virtual components superimposed simultaneously within given PSNADS have been found. The PSNADS can be used as a natural framework to treat the quantum coherence directly in the terms of quantum states. The quantum coherence as a phase property of the wave function can be consistently understood within the dynamics-statistical interpretation of quantum mechanics.